\documentclass[prd,onecolumn,notitlepage,eqsecnum,nofootinbib,floatfix]{revtex4-1}
\usepackage{amssymb}
\usepackage{amsmath}
\usepackage{amsfonts} 
\usepackage{bm}
\usepackage{graphicx}
\DeclareSymbolFont{EulerScript}{U}{eus}{m}{n}
\SetSymbolFont{EulerScript}{bold}{U}{eus}{b}{n}
\DeclareSymbolFontAlphabet\scrpt{EulerScript}
\allowdisplaybreaks
\newcommand{\stf}[1]{{\langle #1 \rangle}}
\newcommand{\gothj}{\mathfrak{j}} 
\newcommand{\gothp}{\mathfrak{p}} 
\newcommand{\e}[1]{{\times 10^{#1}}}   
\begin{document}
\title{Gravitomagnetic  Love tensor of a slowly rotating body: post-Newtonian theory}  
\author{Eric Poisson}  
\affiliation{Department of Physics, University of Guelph, Guelph,
  Ontario, N1G 2W1, Canada} 
\date{August 11, 2020} 
\begin{abstract} 
The gravitomagnetic tidal Love number of a slowly rotating body was calculated previously under the assumption that the velocity perturbation created by the tidal field consists of an induction piece proportional to the vector potential, and a rotational piece that scales with $\Omega$, the body's angular velocity. The second part of this assumption is wrong: the rotational piece of the velocity perturbation scales in fact like $\Omega^0 = 1$. The previous calculations are therefore incorrect, and the purpose of this paper is to repair the mistake. To keep the technical difficulties to a minimum, the treatment here is restricted to a post-Newtonian expansion carried out to leading order --- previous calculations of the gravitomagnetic Love number were performed in full general relativity. On the other hand, the computation presented here is not restricted to a stationary tidal field. I show that the correct scaling of the velocity perturbation with $\Omega$ leads to the promotion of the Love number to a {\it Love tensor}\ $k_{jk}^{\ \ pq}$, a four-index object that relates the body's current quadrupole moment $S_{jk}$ to the gravitomagnetic tidal moment ${\cal B}_{pq}$. The tensorial nature of this quantity has to do with the fact that each $e^{im\phi}$ piece of the tidal force gives rise to an $m$-specific velocity perturbation, and therefore to a Love number that depends on $m$. The collection of these $m$-specific Love numbers makes up the Love tensor $k_{jk}^{\ \ pq}$. 
\end{abstract} 
\maketitle

\section{Introduction} 
\label{sec:intro} 

The observation, by Flanagan and Hinderer \cite{flanagan-hinderer:08}, that the tidal polarizability of a neutron star can be measured in gravitational waves from coalescing compact binaries triggered a sustained effort to determine just how well this measurement can be made, and what information it can deliver on the equation of state of neutron-star matter. The vast literature devoted to this important topic is the subject of a recent review \cite{chatziioannou:20}, and the reader is referred to it for a complete set of references. A measurement of the tidal polarizability was attempted with GW170817 \cite{GW170817:18}, and it delivered an upper bound of astrophysical significance, providing a useful constraint on the equation of state. Future measurements will tighten this bound, and constraints on the equation of state will be further refined with NICER determinations of the mass and radius of X-ray pulsars (see, for example, Ref.~\cite{miller-etal:19} for recent results on PSR J0030+0451).   

Most of the work on tidal polarizability focuses on the deformation of a neutron star created by the gravitoelectric, Newtonian-like field created by the companion body, which dominates the description of the tidal interaction in a post-Newtonian expansion. Most of this work takes the neutron star to be nonrotating. The tidal interaction, however, becomes richer when one includes gravitomagnetic effects produced by the mass currents associated with the companion's orbital motion, and allows the neutron star to be rotating. While the dominant, gravitoelectric tidal deformation can be characterized in terms of a Love number $k_2^{\rm el}$ (with the superscript standing for ``gravitoelectric'', and the subscript standing for $\ell = 2$ or ``quadrupole", the leading term in a multipole expansion of the tidal field), the gravitomagnetic deformation of a nonrotating body requires the introduction of an additional Love number, denoted $k_2^{\rm mag}$ \cite{damour-nagar:09, binnington-poisson:09}. When the neutron star is rotating, the coupling between rotation and tidal field modifies the deformation and requires the introduction of additional, rotational-tidal Love numbers \cite{landry-poisson:15a, pani-etal:15a, pani-etal:15b}. In principle, all these measures of tidal deformation leave an imprint in the phasing of gravitational waves; when determined in observations, they reveal complementary information regarding the internal structure of a neutron star.

The gravitomagnetic Love number $k_2^{\rm mag}$ was calculated by Landry and Poisson \cite{landry-poisson:15b} 
for a nonrotating body in an irrotational fluid state, and this calculation was believed to apply unchanged to a slowly rotating body. The rotational-tidal Love numbers were calculated by Pani, Gualtieri and Ferrari \cite{pani-etal:15b}, 
Landry \cite{landry:17}, and Gagnon-Bischoff {\it et al.}\ \cite{gagnon-bischoff:18} for relativistic polytropes and realistic models of neutron stars. In addition, the coupling between rotation and gravitomagnetic tidal field was shown by Landry and Poisson \cite{landry-poisson:15c} to produce a velocity perturbation that grows linearly with time when the field is idealized as stationary; this fully relativistic prediction was confirmed in a post-Newtonian setting by Poisson and Dou\c{c}ot \cite{poisson-doucot:17}. 

Unfortunately, this entire body of work is wrong: $k_2^{\rm mag}$ is not the same for rotating and nonrotating bodies, the rotational-tidal Love numbers were not computed correctly, and the prediction that a stationary gravitomagnetic tidal field creates a dynamical response in the stellar fluid is incorrect. The reason for this is that all these calculations are based on an invalid assumption ({\it mea culpa}!) regarding the scaling of the velocity perturbation $\delta v^a$ with the body's angular velocity $\Omega$.

To diagnose the problem, let us consider a rotating fluid body subjected to a gravitomagnetic tidal force. The fluid is assumed to have a barotropic equation of state, and its physics is described within the framework of Newtonian mechanics. One of the governing equations is the perturbed Euler equation, 
\begin{equation}
\partial_t \delta v_a + \delta v^b \nabla_b v_a + v^b \nabla_b \delta v_a + \nabla_a (\delta h + \delta U)
= \frac{4}{c^2} \bigl[ \partial_t U_a + v^b (\nabla_b U_a - \nabla_a U_b) \bigr],
\label{euler} 
\end{equation}
where $v^a$ is the unperturbed velocity, $\delta v^a$ its perturbation, $\delta h$ the perturbation of the specific enthalpy, $\delta U$ the perturbation of the Newtonian gravitational potential, $U_a$ the post-Newtonian vector potential responsible for the tidal force, and $\nabla_a$ the (covariant) derivative operator. In the absence of rotation we have that $v^a = 0$, and the solution to Eq.~(\ref{euler}) is $\delta v_a = (4/c^2) U_a$ with $\delta h = \delta U = 0$; the velocity field is the result of gravitomagnetic induction \cite{shapiro:96, favata:06}. The assumption made in the works described previously is that in the case of a rotating body, the velocity perturbation would change to
\begin{equation}
\delta v_a = \frac{4}{c^2} U_a + w_a, \qquad w_a = O(\Omega), 
\label{ansatz} 
\end{equation}
with the additional term $w_a$ assumed to be proportional to $\Omega$, to leading order in an expansion in powers of the angular velocity. Making the substitution in Eq.~(\ref{euler}) and linearizing with respect to $\Omega$, we obtain 
\begin{equation}
\partial_t w_a + \nabla_a (\delta h - \delta U)
= -\frac{4}{c^2} \bigl( v^b \nabla_a U_b + U^b \nabla_b v_a \bigr),
\label{w_eq} 
\end{equation}
and see immediately how a time-independent tidal field can give rise to a $w_a$ that grows linearly with time. The ansatz of Eq.~(\ref{ansatz}) also implies, upon further analysis, that the gravitomagnetic Love number $k_2^{\rm mag}$ is not changed at order $\Omega^0$, and that the rotational-tidal Love numbers have to do with $w_a$, which satisfies the simple equation (\ref{w_eq}). While this discussion is restricted to a post-Newtonian description of the phenomenon, it can be extended to a fully relativistic treatment. The key element behind the previous results is the assumption of Eq.~(\ref{ansatz}), or its relativistic generalization. 

The expectation that $w_a = O(\Omega)$ is wrong ({\it mea maxima culpa}!). It cannot be justified on the basis of Eq.~(\ref{euler}), which can easily accommodate solutions with $w_a = O(1)$. In fact, a careful analysis of the fluid equations by Lockitch and Friedman \cite{lockitch-friedman:99} (see also Refs.~\cite{lockitch-andersson-friedman:00, lockitch-friedman-andersson:03} for a relativistic generalization) revealed the existence of a class of normal modes that couple to a gravitomagnetic tidal force to produce a $w_a$ that scales as $\Omega^0 = 1$ instead of $\Omega$ \cite{poisson:20a}. These modes are known as {\it inertial modes}, because the restoring force is provided by the fluid's rotation (the Coriolis force, when viewed in the body's corotating frame). The existence of inertial modes, therefore, invalidates the assumption that $w_a = O(\Omega)$. Instead we have $w_a = O(1)$, and this revised scaling comes with a number of consequences. First, the fact that $\delta v_a$ is modified at order $\Omega^0$ means that the gravitomagnetic Love number $k_2^{\rm mag}$ cannot be the same for rotating and nonrotating bodies (everything else being equal). Second, because the rotational-tidal Love numbers arise from terms of order $\Omega$ in the fluid perturbation, their calculation require a $\delta v_a$ that's expanded through order $\Omega$; this was not done correctly in previous calculations. And third, there is no more ground to expect a stationary tidal field to create a dynamical fluid response. 

My purpose with this paper is to repair the calculation of $k_2^{\rm mag}$ for a slowly rotating body. To keep the technical difficulties to a minimum, I shall restrict my analysis to a leading-order, post-Newtonian treatment; promotion to full general relativity will await future work. On the other hand, unlike in previous works, the computations presented here allow the tidal field to depend on time. A correct calculation of the rotational-tidal Love numbers will not be attempted here; this also will be left for future work.

I set the stage in Sec.~\ref{sec:nonrotating} with a refresher on how $k_2^{\rm mag}$ is defined for a nonrotating body, as the proportionality between a gravitomagnetic tidal moment ${\cal B}_{jk}$ and the resulting current quadrupole moment $S_{jk}$ of the matter distribution. In Sec.~\ref{sec:rotating}, I identify what is required in a generalization to a slowly rotating body, given that the velocity perturbation continues to scale as $\Omega^0$. I introduce the notion that the Love number must be replaced by a {\it Love tensor} $k_{jk}^{\ \ pq}$, and I specify the goals of the calculation: to obtain the Love tensor to leading order in a post-Newtonian expansion of the gravitational field, and to leading order in an expansion in powers of $\Omega$.

The calculation begins in Sec.~\ref{sec:pN} with a review of the relevant post-Newtonian field equations. I show that the body's gravitomagnetic response can be described in terms of two contributions to the vector potential $U_j$, one coming directly from the tidal field, the other coming indirectly through the velocity perturbation. In Sec.~\ref{sec:field}, I calculate the field contribution to the vector potential, and obtain the corresponding contribution to the current quadrupole moment. The calculation of the matter contribution occupies the following three sections. In Sec.~\ref{sec:decomposition}, I decompose the tidal tensor ${\cal B}_{jk}$ into $m$-components, where $m$ is the azimuthal integer that characterizes the $e^{im\phi}$ behavior of each component of the tidal field. I introduce the velocity perturbation $\delta v_j$ in Sec.~\ref{sec:tidal_force}, and show that it admits a complicated decomposition in radial, polar, and axial spherical harmonics. I calculate the associated vector potential in Sec.~\ref{sec:matter}, determine the matter contribution to the current quadrupole moment, and obtain Love numbers for each implicated value of $m$.

This dependence of the Love numbers on $m$ implies that the body's response must be described by a Love tensor instead of a single number. I collect the results in Sec.~\ref{sec:tensor}, and construct the complete Love tensor $k_{jk}^{\ \ pq}$ by combining the field and matter contributions to the current quadrupole moment. This concludes the paper. 

Some incidental matters are relegated to Appendices. I examine the $\Omega \to 0$ limit of the velocity perturbation of Sec.~\ref{sec:tidal_force} in Appendix~\ref{sec:norotation}, and show that it reduces to the expected $\delta v_j = (4/c^2) U_j$. I review the multipole expansion of the vector potential in Appendix~\ref{sec:anapole}. I show that it involves the angular-momentum vector $S^j$ at leading order, and that the next order features an anapole moment $K^j$ in addition to the current quadrupole moment $S^{jk}$; the anapole term is a pure gradient that can always be eliminated with a gauge transformation. In Appendix~\ref{sec:geometric}, I provide an alternative expression for the Love tensor of Sec.~\ref{sec:tensor}, in which it is expressed in terms of geometric quantities intrinsic to the situation. 

The technical developments in the paper rely heavily on Ref.~\cite{poisson:20a}, which develops a lot of the required infrastructure. I use indices $jkp\ldots$ in all equations formulated in Cartesian coordinates, and indices $abc\ldots$ in covariant equations, or in equations written in spherical polar coordinates.  

\section{Gravitomagnetic response of a nonrotating body} 
\label{sec:nonrotating} 

A fully relativistic definition of the gravitomagnetic Love number $k_2^{\rm mag}$ of a nonrotating body is provided, for example, in Ref.~\cite{landry-poisson:15a}. We review this definition here. 

A body of mass $M$ and radius $R$ is immersed in a time-varying gravitomagnetic tidal field characterized by the quadrupole-moment tensor ${\cal B}_{jk}(t)$. In suitable Lorentzian-like coordinates $(ct,x^j)$ defined in the body's local asymptotic rest-frame, the time-space components of the metric tensor are given by 
\begin{equation} 
g_{0j} = \frac{2}{3c^3} \epsilon_{jkp} {\cal B}^p_{\ q} x^k x^q \biggl[ (1 + \cdots) 
- 6 \frac{GM R^4}{c^2 r^5} k_2^{\rm mag} (1 + \cdots) \biggr], 
\label{g0j} 
\end{equation} 
where $\epsilon_{jkp}$ is the antisymmetric permutation symbol, $r^2 := \delta_{jk} x^j x^k$, and the ellipsis denotes relativistic corrections of order $GM/(c^2 r)$ and higher. The first term in $g_{0j}$ represents the external tidal field; the second term is the body's response, measured by the gravitomagnetic tidal Love number. 

We shall be interested in a post-Newtonian approximation of the metric, and for this purpose it is convenient to introduce the vector potential $U_j := -\frac{1}{4} c^3 g_{0j}$. We decompose it as 
\begin{equation} 
U_j = U_j^{\rm tidal} + U_j^{\rm body}, 
\label{Uj_decomp1} 
\end{equation} 
with
\begin{equation} 
U_j^{\rm tidal} = -\frac{1}{6} \epsilon_{jkp} {\cal B}^p_{\ q} x^k x^q + O(c^{-2})
\label{Uj_tidal} 
\end{equation} 
describing the tidal field, and
\begin{equation} 
U_j^{\rm body} = \frac{GM R^4}{c^2} k_2^{\rm mag} 
\epsilon_{jkp} {\cal B}^p_{\ q} \frac{x^k x^q}{r^5} + O(c^{-4})
\end{equation} 
representing the body's response. This can be measured in terms of a current quadrupole moment $S_{jk}$, such that 
\begin{equation}
U_j^{\rm body} = -G  \epsilon_{jkp} S^p_{\ q} \frac{x^k x^q}{r^5}. 
\label{Uj_body} 
\end{equation} 
Comparison with the previous expression reveals the relationship between $S_{jk}$ and ${\cal B}_{jk}$; we have 
\begin{equation} 
S_{jk} = -\frac{M R^4}{c^2} k_2^{\rm mag} {\cal B}_{jk} + O(c^{-4}). 
\label{S_vs_B} 
\end{equation} 
We shall take this as the official definition of the gravitomagnetic Love number, to leading order in a post-Newtonian treatment of the tidal interaction. 

\section{Gravitomagnetic response of a slowly rotating body} 
\label{sec:rotating} 

The gravitomagnetic tidal response of a slowly rotating body can still be described in terms of a current quadrupole moment $S_{jk}$. We still decompose the vector potential as in Eq.~(\ref{Uj_decomp1}), with the tidal field of Eq.~(\ref{Uj_tidal}) and the body's response of Eq.~(\ref{Uj_body}). But the relationship between $S_{jk}$ and ${\cal B}_{jk}$ is now a lot more complicated. We shall see in Sec.~\ref{sec:tensor} that it becomes 
\begin{equation} 
S_{jk}(t) = -\frac{MR^4}{c^2} \int_{-\infty}^\infty 
\tilde{k}_{jk}^{\ \ pq}(\omega) \tilde{\cal B}_{pq}(\omega)\, e^{-i\omega t}\, d\omega, 
\label{S_vs_B_convolution} 
\end{equation} 
a Fourier integral implicating a gravitomagnetic Love tensor $\tilde{k}_{jk}^{\ \ pq}(\omega)$ and the frequency-domain tidal moment $\tilde{\cal B}_{pq}(\omega)$. We omit the labels ``mag'' and ``2'' on the Love tensor, as the context is now clearly identified: we focus our attention on the gravitomagnetic response of a slowly rotating body at the leading, quadrupole order in a multipole expansion of the tidal interaction. The Love tensor will be given a concrete expression in Sec.~\ref{sec:tensor}.  

We shall see that the current quadrupole moment can be decomposed as 
\begin{equation} 
S_{jk} = S^{\rm M}_{jk} + S^{\rm F}_{jk}, 
\label{S_decomp} 
\end{equation} 
where $S^{\rm M}_{jk}$ is a contribution from the matter distribution inside the body, while $S^{\rm F}_{jk}$ is a contribution from the gravitational field. We shall decompose the Love tensor in a similar manner, according to 
\begin{equation} 
k_{jk}^{\ \ pq} = k_{jk}^{{\rm M}\, pq} + k_{jk}^{{\rm F}\, pq}. 
\end{equation} 
We aim to calculate the Love tensor to leading order in simultaneous expansions in powers of $c^{-2}$ (a post-Newtonian expansion) and $\Omega$ (a slow-rotation expansion). We shall see that these leading orders are 
\begin{equation} 
k_{jk}^{\ \ pq} = O(c^0, \Omega^0). 
\label{love_orders} 
\end{equation} 
A key point is that {\it the Love tensor differs from the Love number of a nonrotating body at zeroth-order in an expansion in powers of} $\Omega$; the gravitomagnetic tidal response of a slowly rotating body is radically different from the response of a nonrotating body.

The way in which the Love tensor $k_{jk}^{\ \ pq}$ collapses to a Love number $k$ when $\Omega \to 0$ is subtle. We shall elucidate this limit in the remaining sections of the paper. 

\section{post-Newtonian field equations} 
\label{sec:pN} 

We now proceed with the calculation. The field equations of post-Newtonian gravity are summarized in Sec.~7.1 of
Ref.~\cite{poisson-will:14}. With $-4 c^{-3} U_j = g_{0j} = -h^{0j}[1 + O(c^{-2})]$ and $s^{j} := c^{-1} \tau^{0j}$ (in the notation used there), we have that the vector potential satisfies the Poisson equation 
\begin{equation} 
\nabla^2 U_j = -4\pi G s_j.   
\label{Uj_poisson} 
\end{equation} 
An explicit expression for the current density is provided in Exercise 8.4 of Ref.~\cite{poisson-will:14}; we have that 
\begin{equation} 
s_j = \rho v^{\rm tot}_j [1 + O(c^{-2})] + \frac{1}{\pi G c^2} (\partial_j U^k - \partial^k U_j) \partial_k U 
+ O(c^{-4}),  
\label{sj} 
\end{equation} 
where $\rho$ is the mass density, $v^{\rm tot}_j$ the body's velocity field, and $U$ the Newtonian gravitational potential. The velocity is decomposed as 
\begin{equation} 
v^{\rm tot}_j = v_j + \delta v_j, \qquad  
v_j = \epsilon_{jkp} \Omega^k x^p, 
\label{vj} 
\end{equation} 
with $v_j$ representing the rigid rotation of the unperturbed body ($\Omega^k$ is the angular-velocity vector), while $\delta v_j$ is the perturbation produced by the tidal interaction. 

The vector potential appears on both sides of Eq.~(\ref{Uj_poisson}), and the usual strategy to deal with this complication is to iterate. In a first iteration we aim to obtain the potential at order $c^{0}$. We therefore ignore the $c^{-2}$ term in Eq.~(\ref{sj}), and incorporate only the rotational velocity in $v^{\rm tot}_j$; we neglect the velocity perturbation because it scales as $c^{-2}$, being the result of a post-Newtonian tidal interaction. We obtain   
\begin{equation} 
U_j = U^{\rm rot}_j + U^{\rm tidal}_j + O(c^{-2}),  
\end{equation} 
where the second term is the tidal potential of Eq.~(\ref{Uj_tidal}) --- a solution to Laplace's equation --- while
\begin{equation} 
U^{\rm rot}_j = \frac{1}{2} G \epsilon_{jkp} S^k \frac{x^p}{r^3} 
\end{equation} 
is the rotational piece. The vector 
\begin{equation} 
S^j := I^{jk} \Omega_k, \qquad 
I^{jk} := \int \rho\bigl( r^2 \delta^{jk} - x^j x^k \bigr)\, dV 
\end{equation} 
is the spin angular momentum, expressed here in terms of the angular-velocity vector and the body's moment of inertia.  

In a second iteration of the field equations we incorporate $\delta v_j$ and insert the old $U_j$ within the current density $s_j$ in Eq.~(\ref{sj}); and we integrate Eq.~(\ref{Uj_poisson}) again to obtain an improved version of the vector potential, accurate through order $c^{-2}$. Our goal is to obtain the corrections to $U_j$ to leading order in an expansion in powers of $\Omega$. Because $\delta v_j$ scales as $\Omega^0$, this leading order is $\Omega^0$, and we allow ourselves to neglect all contributions that scale with a higher power of $\Omega$. Thus, while in principle we should account for the corrections of order $c^{-2}$ that come with $v_j$ in $s_j$, we see that these would merely alter the rotational piece of the vector potential, which scales as $\Omega$. Similarly, we may discard the rotational piece of the potential when we make the substitution within $s_j$, because it would also give rise to a $O(\Omega)$ correction to the iterated vector potential.

With all this understood, the second iteration produces 
\begin{equation} 
U_j = U^{\rm rot}_j + U^{\rm tidal}_j 
+ U^{\rm M}_j + U^{\rm F}_j + O(c^{-4}),  
\label{Uj_second} 
\end{equation} 
where the new contributions to the vector potential are solutions to 
\begin{equation} 
\nabla^2 U_j^{\rm M} = -4\pi G \rho \delta v_j 
\label{UM_poisson} 
\end{equation} 
and 
\begin{equation} 
\nabla^2 U^{\rm F}_j = -4\pi s^{\rm F}_j, 
\label{UF_poisson} 
\end{equation} 
with 
\begin{equation} 
s^{\rm F}_j := \frac{1}{\pi c^2} \bigl( \partial_j U^k_{\rm tidal} - \partial_k U^{\rm tidal}_j \bigr) \partial_k U. 
\label{sj_F} 
\end{equation}  
The third term in Eq.~(\ref{Uj_second}) is a matter contribution to the vector potential that is associated with the velocity perturbation, while the fourth is a contribution that comes directly from the tidal field. Both scale as $O(c^{-2}, \Omega^0)$, the orders required in a leading-order calculation of the Love tensor, as expressed by Eq.~(\ref{love_orders}). 

\section{Field contribution to the vector potential} 
\label{sec:field} 

The solution to Eq.~(\ref{UF_poisson}) for $U^{\rm F}_j$ was previously constructed (for a tidal field of arbitrary multipole order) in Ref.~\cite{landry-poisson:15b}. We review this calculation here, restricted to the specific case of a quadrupolar field.  

We insert the tidal potential of Eq.~(\ref{Uj_tidal}) within Eq.~(\ref{sj_F}), along with $\partial_k U = -G m(r) r_k/r^2$ for the gradient of the Newtonian potential; $m(r)$ is the body's internal mass function, and $r_j = \partial_j r$ is the unit radial vector. We obtain 
\begin{equation} 
s_j^{\rm F} = -\frac{Gm}{2\pi c^2 r} \epsilon_{jkp} {\cal B}^p_{\ q} r^\stf{kq}, 
\end{equation} 
where $r^\stf{kq} := r^k r^q - \frac{1}{3} \delta^{kq}$. The solution to Eq.~(\ref{UF_poisson}) is 
\begin{equation} 
U^{\rm F}_j (\bm{x}) = \int \frac{s^{\rm F}_j(\bm{x'})}{|\bm{x}-\bm{x'}|}\, dV', 
\end{equation} 
where $dV'$ is an element of volume centered at $\bm{x'}$. To evaluate the integral we make use of the addition theorem for spherical harmonics, as well as Eq.~(1.171) of Ref.~\cite{poisson-will:14} to dispose of the angular integrations. We obtain 
\begin{equation} 
U^{\rm F}_j = -\frac{2G}{5c^2} \epsilon_{jkp} {\cal B}^p_{\ q} r^\stf{kq}\, J 
\end{equation} 
with 
\begin{equation} 
J := \int_0^{\cal R} m(r') r' \frac{r_<^2}{r_>^3}\, dr', 
\end{equation} 
where $r_< := \mbox{min}(r,r')$ and $r_> := \mbox{max}(r,r')$. The radial integral is regularized with an arbitrary cutoff radius ${\cal R}$, imagined to be large compared with $R$, but small compared with the scale of variation of the external tidal field. 

We are interested in $U^{\rm F}_j$ evaluated outside the body (where $r > R$). To compute $J$ we break up the integration domain into three segments, the first going from $0$ to $R$, the second from $R$ to $r$, and the third from $r$ to ${\cal R}$. In the first segment, $m$ is a function of $r'$, $r_< = r'$, and $r_> = r$. In the second segment, $m$ becomes the constant $M$, and we still have $r_< = r'$ and $r_> = r$. In the third segment, $m = M$, but we now have $r_< = r$, and $r_> = r'$. To evaluate the contribution from the first segment we integrate by parts, making use of $dm/dr = 4\pi r^2\rho$. The end result is 
\begin{equation} 
J = -\frac{\pi}{r^3} \int_0^R \rho\, r^6\, dr + \frac{5}{4} Mr - \frac{Mr^2}{\cal R}. 
\end{equation} 
The term proportional to $Mr$ gives rise to a correction of order $GM/(c^2 r)$ to the tidal potential; we shall not be interested in this correction, and we henceforth discard it. The term involving ${\cal R}$ represents a shift of ${\cal B}_{jk}$ by a quantity of order $GM/(c^2 {\cal R})$; such a shift is meaningless, and we also eliminate this term. 

We have arrived at  
\begin{equation} 
U^{\rm F}_j = -G \epsilon_{jkp} S^{\ p}_{{\rm F}\, q} \frac{x^k x^q}{r^5}, 
\label{Uj_F_final} 
\end{equation} 
with 
\begin{equation} 
S^{\rm F}_{jk} = -\frac{MR^4}{c^2} k^{\rm F} {\cal B}_{jk}
\label{Sjk_F} 
\end{equation} 
and 
\begin{equation} 
k^{\rm F} := \frac{3}{10} \int_0^1 \hat{\rho}\, \hat{r}^6\, d\hat{r}, 
\label{k_F} 
\end{equation} 
where $\hat{\rho} := 4\pi R^3 \rho/(3M)$ is a dimensionless density function, and $\hat{r} := r/R$ a dimensionless radius. This expression for the field contribution to the vector potential agrees with Eq.~(6.20) of Ref.~\cite{landry-poisson:15b} when $\ell = 2$ and $\lambda = 0$. For this piece of the potential, the Love quantity that relates the current quadrupole moment $S_{jk}$ to the tidal quadrupole moment ${\cal B}_{jk}$ is actually a scalar --- a Love number. This can be promoted to a tensor by writing 
\begin{equation} 
k^{{\rm F}\, pq}_{jk} := k^{\rm F}\, \delta_j^{\ p} \delta_k^{\ q}. 
\label{kF_tensor} 
\end{equation} 
 
\section{Decomposition of the tidal quadrupole moment} 
\label{sec:decomposition} 

In the following sections we will endeavor to integrate Eq.~(\ref{UM_poisson}) for the matter contribution to the vector potential. To aid this calculation we introduce the Fourier transform of the tidal tensor ${\cal B}_{jk}(t)$, as well as a decomposition into a tensorial basis that projects out each one of its $m$-components, with $m$ the azimuthal integer that characterizes the $e^{im\phi}$ behavior of each piece of the tidal potential.  

The Fourier decomposition of the tidal moment is given by 
\begin{equation} 
{\cal B}_{jk}(t) = \int_{-\infty}^\infty \tilde{\cal B}_{jk}(\omega) e^{-i\omega t}\, d\omega, 
\end{equation} 
where $\omega$ is the frequency, measured in the body's inertial frame (and not in the corotating frame, as is sometimes done). Because the tidal moment is real, its Fourier transform satisfies 
\begin{equation} 
\tilde{\cal B}_{jk}(-\omega) = \tilde{\cal B}^*_{jk}(\omega),
\end{equation} 
where an asterisk indicates complex conjugation. In our subsequent developments we shall also use an overbar to denote complex conjugation. For conceptual convenience we take $\tilde{\cal B}_{jk}(\omega)$ to be peaked at frequencies $\omega$ that are comparable to $\Omega$, so that $\omega/\Omega$ can formally be taken to be of order unity.  

The decomposition into $m$-components is accomplished with the help of the symmetric-tracefree tensors (see Box~1.5 of Ref.~\cite{poisson-will:14})  
\begin{equation} 
({\scrpt Y}_2^{\pm 2})^{jk} = \frac{1}{8} \sqrt{\frac{30}{\pi}} \left( 
\begin{array}{ccc} 
1 & \mp i & 0 \\ 
\mp i & -1 & 0 \\ 
0 & 0 & 0 
\end{array} \right), \qquad 
({\scrpt Y}_2^{\pm 1})^{jk} = \mp \frac{1}{8} \sqrt{\frac{30}{\pi}} \left( 
\begin{array}{ccc} 
0 & 0 & 1 \\ 
0 & 0 & \mp i \\ 
1 & \mp i & 0 
\end{array} \right), \qquad 
({\scrpt Y}_2^{0})^{jk} = -\frac{1}{4} \sqrt{\frac{5}{\pi}} \left( 
\begin{array}{ccc} 
1 & 0 & 0 \\ 
0 & 1 & 0 \\ 
0 & 0 & -2 
\end{array} \right), 
\label{STFtensors} 
\end{equation}   
which satisfy $({\scrpt Y}_2^{-m})^{jk} = (-1)^m (\bar{\scrpt Y}_2^{m})^{jk}$. They are defined so that the spherical harmonics of degree $\ell = 2$ can be expressed as [Eq.~(1.167) of Ref.~\cite{poisson-will:14}]  
\begin{equation} 
Y_2^m(\theta,\phi) =  (\bar{\scrpt Y}_2^{m})_{jk}\, r^j r^k, 
\end{equation} 
where $r^j = (\sin\theta\cos\phi, \sin\theta\sin\phi, \cos\theta)$ is the radial unit vector. It is useful to record the completeness relations
\begin{equation}
\frac{8\pi}{15} (\bar{\scrpt Y}_2^{m})^{jk} ({\scrpt Y}_2^{m'})_{jk} = \delta_{mm'}
\end{equation}
and
\begin{equation}
\frac{8\pi}{15} \sum_{m=-2}^2 (\bar{\scrpt Y}_2^{m})^{jk} ({\scrpt Y}_2^{m})_{pq}
= \frac{1}{2} \delta^j_{\ p} \delta^k_{\ q} + \frac{1}{2} \delta^j_{\ q} \delta^k_{\ p}
- \frac{1}{3} \delta^{jk} \delta_{pq}. 
\label{completeness} 
\end{equation}
The operator on the right-hand side of Eq.~(\ref{completeness}) takes any tensor $A^{pq}$ and returns its symmetric-tracefree part $A^\stf{jk}$; the same is true, therefore, of the operator on the left-hand side. 

The decomposition of the tidal moment is given by 
\begin{equation} 
\tilde{\cal B}_{jk} = \sum_{m=-2}^2 \tilde{\cal B}^m_{jk},
\label{B_decomposition1} 
\end{equation}
with
\begin{equation}
\tilde{\cal B}^m_{jk} :=\tilde{\cal B}^m  (\bar{\scrpt Y}_2^{m})_{jk}
\label{B_decomposition2} 
\end{equation} 
and 
\begin{equation} 
\tilde{\cal B}^m := \frac{8\pi}{15} ({\scrpt Y}_2^{m})^{jk}\, \tilde{\cal B}_{jk}. 
\label{B_projections} 
\end{equation} 
The projections satisfy the reality conditions 
\begin{equation} 
\tilde{\cal B}^{-m}(\omega) = (-1)^m \bigl[ \tilde{\cal B}^m(-\omega) \bigr]^*. 
\label{reality} 
\end{equation} 

As a specific example of a gravitomagnetic tidal moment, we take the field produced by a companion body of mass $M'$ moving on a circular orbit of radius $p$ and angular velocity $\varpi$, with $\varpi^2 = G(M+M')/p^3$. The companion's position with respect to the body's center of mass is in the direction of the unit vector $\bm{n}$, and the normal to the orbital plane points along the unit vector $\bm{l}$. We give the orbit a generic orientation, so that it possesses an inclination angle $\iota$ with respect to the body's equatorial plane. The orbital vectors are given by [Eqs.~(3.42) and (3.45) of Ref.~\cite{poisson-will:14}]  
\begin{equation}  
\bm{n} = \bigl( \cos\varpi t, \cos\iota\, \sin\varpi t, \sin\iota\, \sin\varpi t \bigr), \qquad  
\bm{l} = \bigl(0, -\sin\iota, \cos\iota \bigr), 
\label{orbit_vectors}
\end{equation} 
The tidal moment is (see, for example, Ref.~\cite{taylor-poisson:08}) 
\begin{equation} 
{\cal B}_{jk} = \frac{3GM'v'}{p^3} \bigl( l_j n_k + n_k l_k \bigr), 
\end{equation} 
where $v' = p\varpi$ is the orbital velocity. Inserting these in Eq.~(\ref{B_projections}) and evaluating the Fourier transforms, we arrive at
\begin{subequations} 
\label{B_binary} 
\begin{align} 
\tilde{\cal B}^{m=2} &= i\frac{\sqrt{30\pi}}{5} \frac{GM' v'}{p^3} \Bigl[ 
\sin\iota(1 + \cos\iota)\, \delta(\omega - \varpi) 
+ \sin\iota(1 - \cos\iota)\, \delta(\omega + \varpi) \Bigr], \\ 
\tilde{\cal B}^{m=1} &= \frac{\sqrt{30\pi}}{5} \frac{GM' v'}{p^3} \Bigl[ 
 (1+\cos\iota)(1-2\cos\iota)\, \delta(\omega - \varpi) 
- (1-\cos\iota)(1+2\cos\iota)\, \delta(\omega + \varpi) \Bigr], \\ 
\tilde{\cal B}^{m=0} &= i\frac{6\sqrt{5\pi}}{5} \frac{GM' v'}{p^3} \Bigl[ 
\sin\iota\cos\iota\, \delta(\omega - \varpi) 
- \sin\iota\cos\iota\, \delta(\omega + \varpi) \Bigr]. 
\end{align} 
\end{subequations} 
The projections with $m < 0$ can be obtained with the help of Eq.~(\ref{reality}). 

\section{Velocity perturbation}
\label{sec:tidal_force} 

The calculation of $\delta v_j$ for a slowly rotating body subjected to a generic gravitomagnetic tidal field is virtually identical to the one presented in Ref.~\cite{poisson:20a}, which applied specifically to a field produced by a companion body moving on a circular orbit. We sketch this calculation here; a wealth of additional details can be found in the earlier reference. 

The body is subjected to a tidal force density produced by the vector potential
\begin{equation}
\tilde{U}_j = -\frac{1}{6} \epsilon_{jkp} \tilde{\cal B}^p_{\ q} x^k x^q;
\end{equation}
this is the Fourier transform of Eq.~(\ref{Uj_tidal}), and we omit the label ``tidal'' to avoid cluttering the notation. The force density is denoted $\rho f_j$, and it is given by
\begin{equation}
\tilde{f}_j = \frac{4}{c^2} \bigl[ -i\omega \tilde{U}_j
+ v^k (\partial_k \tilde{U}_j - \partial_j \tilde{U}_k) \bigr],
\end{equation}
where $v_j := \epsilon_{jkp} \Omega^k x^p$ is the body's rotational velocity.

To calculate $\delta v_j$ from the equations of fluid dynamics, it is sufficient to consider the curl of the force density, which we denote
\begin{equation}
\tilde{q}^j = \epsilon^{jkp} \partial_k \tilde{f}_p,
\end{equation}
and which we decompose according to
\begin{equation}
\tilde{q}^a = \sum_{m=-2}^2 \tilde{q}^a_m e^{im\phi}.
\label{q_decomp} 
\end{equation}
The switch from a vector index $j$ to an index $a$ is meant to indicate that the components of $\tilde{q}^a$ are calculated in spherical coordinates $(r,\theta,\phi)$. The computation makes use of the decomposition of the tidal tensor into $m$-components, as given by Eq.~(\ref{B_decomposition1}). We obtain
\begin{subequations}
\label{q_comps_m2} 
\begin{align}
\tilde{q}^r_{\pm 2} &= \pm\frac{i}{4} \sqrt{\frac{30}{\pi}} \frac{\tilde{\cal B}^{\pm 2}}{c^2}
(2\Omega \mp \omega) r \sin^2\theta, \\
\tilde{q}^\theta_{\pm 2} &= \pm\frac{i}{4} \sqrt{\frac{30}{\pi}} \frac{\tilde{\cal B}^{\pm 2}}{c^2}
(2\Omega \mp \omega) \sin\theta\cos\theta, \\
\tilde{q}^\phi_{\pm 2} &= -\frac{1}{4} \sqrt{\frac{30}{\pi}} \frac{\tilde{\cal B}^{\pm 2}}{c^2}
(2\Omega \mp \omega),
\end{align}
\end{subequations}
\label{q_comps_m1} 
\begin{subequations}
\begin{align}
\tilde{q}^r_{\pm 1} &= -\frac{i}{2} \sqrt{\frac{30}{\pi}} \frac{\tilde{\cal B}^{\pm 1}}{c^2}
(\Omega \mp \omega) r \sin\theta\cos\theta, \\
\tilde{q}^\theta_{\pm 1} &= -\frac{i}{4} \sqrt{\frac{30}{\pi}} \frac{\tilde{\cal B}^{\pm 1}}{c^2}
(\Omega \mp \omega) (2\cos^2\theta - 1), \\
\tilde{q}^\phi_{\pm 1} &= \pm\frac{1}{4} \sqrt{\frac{30}{\pi}} \frac{\tilde{\cal B}^{\pm 1}}{c^2}
(\Omega \mp \omega) \cot\theta, 
\end{align}
\end{subequations}
and
\begin{subequations}
\label{q_comps_m0} 
\begin{align}
\tilde{q}^r_{0} &= -\frac{i}{2} \sqrt{\frac{5}{\pi}} \frac{\tilde{\cal B}^{0}}{c^2}
\omega r (3\cos^2\theta - 1), \\ 
\tilde{q}^\theta_{0} &= \frac{3i}{2} \sqrt{\frac{5}{\pi}} \frac{\tilde{\cal B}^{0}}{c^2}
\omega \sin\theta\cos\theta, \\
\tilde{q}^\phi_{0} &= 0. 
\end{align}
\end{subequations}
These equations can be compared with Eqs.~(3.11), (3.12), and (3.13) of Ref.~\cite{poisson:20a}. The projections $\tilde{q}^a_m(\omega)$ satisfy the same reality conditions as $\tilde{B}^m(\omega)$, as displayed in Eq.~(\ref{reality}). We recall that $\omega$ is taken to be comparable to $\Omega$. 

The velocity perturbation is decomposed as
\begin{equation}
\delta \tilde{v}^a = \sum_{m=-2}^2 \delta \tilde{v}^a_m,
\end{equation}
and each projection is further decomposed into a basis of vectorial harmonics. The first member of the basis consists of the radial harmonics
\begin{equation}
r^a Y_\ell^m,
\label{radial} 
\end{equation}
the union of the usual spherical harmonics $Y_\ell^m$ with the unit radial vector $r^a$. The second member is the set of polar harmonics
\begin{equation}
(Y_\ell^m)^a := \nabla^a Y_\ell^m.
\label{polar} 
\end{equation}
And the third member consists of the axial harmonics
\begin{equation}
(X_\ell^m)^a := \epsilon^{abc} (\nabla_b Y_\ell^m) r_c, 
\label{axial} 
\end{equation}
the cross product between the polar harmonics and the unit radial vector. The decomposition is written as
\begin{subequations}
\label{va_decomp}
\begin{align}
\delta \tilde{v}^a_{\pm 2} &= \frac{\tilde{\cal B}^{\pm 2}}{c^2}
\frac{2 \mp \omega/\Omega}{4 \mp 3\omega/\Omega} r^3 (X_2^{\pm 2})^a,
\label{dv_m2} \\
\delta \tilde{v}^a_{\pm 1} &= \frac{5}{\sqrt{30\pi}} \frac{\tilde{\cal B}^{\pm 1}}{c^2} (1 \mp \omega/\Omega) R^3\Biggl[ - i \sum_{\ell=1}^{\rm odd} \frac{1}{r} A_\ell^{\pm 1}\, r^a Y_\ell^{\pm 1}
- i \sum_{\ell=1}^{\rm odd} B_\ell^{\pm 1}\, (Y_\ell^{\pm 1})^a
\pm \sum_{\ell=2}^{\rm even} C_\ell^{\pm 1}\, (X_\ell^{\pm 1})^a \Biggr], \\
\delta\tilde{v}^a_{0} &= \frac{5}{6\sqrt{5\pi}} \frac{\tilde{\cal B}^{0}}{c^2} (\omega/\Omega) R^3
\Biggl[ - i \sum_{\ell=1}^{\rm odd} \frac{1}{r} A_\ell^{0}\, r^a Y_\ell^{0}
- i \sum_{\ell=1}^{\rm odd} B_\ell^{0}\, (Y_\ell^{0})^a
- \sum_{\ell=2}^{\rm even} C_\ell^{0}\, (X_\ell^{0})^a \Biggr],
\end{align}
\end{subequations}
where the radial functions $A_\ell^m(\omega, r)$, $B_\ell^m(\omega, r)$, and $C_\ell^m(\omega, r)$ are all real and dimensionless; they satisfy
\begin{equation}
A_\ell^{-m}(\omega,r) = (-1)^m A_\ell^m(-\omega,r), \qquad
B_\ell^{-m}(\omega,r) = (-1)^m B_\ell^m(-\omega,r),
\label{AB_reality} 
\end{equation}
as well as
\begin{equation}
C_\ell^{-1}(\omega,r) = -C_\ell^1(-\omega,r), \qquad
C_\ell^{0}(\omega,r) = -C_\ell^0(-\omega,r).
\label{C_reality} 
\end{equation}
These relations imply that
\begin{equation}
\delta \tilde{v}^a_{-m}(\omega) = \bigl[ \delta \tilde{v}^a_m(-\omega) \bigr]^*.
\end{equation} 
Equations (\ref{va_decomp}) can be compared with Eqs.~(4.3) of Ref.~\cite{poisson:20a}. The restrictions on the sums over $\ell$ (even or odd terms only) is justified on the basis of the fluid equations. The strange numerical prefactors, and the various factors of $\pm 1$ and $-i$, are inserted to ensure that the radial functions satisfy the same equations as those listed in Sec.~IV of Ref.~\cite{poisson:20a}; these will not be duplicated here. In principle, the fluid equations produce an infinite set of differential equations for the radial functions, which are all coupled to one another. In practice the system of equations is truncated to include a finite number of radial functions, those with $\ell \leq 6$ in the case of $m=\pm 1$ and $m=0$. A method to integrate these equations numerically is presented in Sec.~V of Ref.~\cite{poisson:20a}, together with a small sampling of the results; again we shall not duplicate this discussion here. In the case of $m = \pm 2$, only one radial function survives the integration of the fluid equations, and it can be obtained analytically --- refer to Eq.~(5.8) of Ref.~\cite{poisson:20a}; this information was already incorporated in Eq.~(\ref{dv_m2}).

The radial functions depend on $\omega$ and $\Omega$ via the dimensionless combination $w := \omega/\Omega$. This, we recall, is taken to be of order unity. With this understanding, Eqs.~(\ref{va_decomp}) indicate that $\delta \tilde{v}^a$ scales as $\Omega^0 = 1$, as was claimed at the beginning of this paper. Relaxing this assumption, it is straightforward to take $\omega$ to be much smaller than $\Omega$ by subjecting Eqs.~(\ref{va_decomp}) to a limit $w \to 0$. The opposite scenario, the nonrotating limit in which $\Omega$ is taken to be much smaller than $\omega$, is more subtle, because $\delta \tilde{v}^a$ appears to be singular when $w \to \infty$. We investigate this issue in Appendix~\ref{sec:norotation}, and conclude that the singularity is only apparent; the nonrotating limit is actually well defined. As a final remark, we note that the limit in which $\omega$ and $\Omega$ both approach zero is ambiguous, because the limiting behavior actually depends on $w$ instead of $\omega$ and $\Omega$ individually. 

\section{Matter contribution to the vector potential}
\label{sec:matter}

With the velocity perturbation now at hand, we may return to the integration of Eq.~(\ref{UM_poisson}) for the matter contribution to the vector potential. We are interested in the solution outside the body, and we provide it as a multipole expansion in powers of $r^{-1}$, with $r$ still denoting the distance to the center of mass. To leading order we have
\begin{equation}
\tilde{U}_j^{\rm M} = -G \epsilon_{jkp} \tilde{S}^{\ p}_{{\rm M}\ q} \frac{x^k x^q}{r^5},  
\end{equation} 
where
\begin{equation}
\tilde{S}_{\rm M}^{jk} = \epsilon^{(j}_{\ \ pq} \int \rho\, x^{k)} x^p \delta \tilde{v}^q\, dV 
\label{SM}
\end{equation}
is the matter contribution to the current quadrupole moment. (The vector potential actually comes with another term featuring an ``anapole'' moment. This term, however, is a pure gradient, and it can always be eliminated by a gauge transformation. We document this in Appendix~\ref{sec:anapole}.)

The calculation of $\tilde{S}_{\rm M}^{jk}$ proceeds as in Sec.~V D of Ref.~\cite{poisson:20a}, and there is no need to duplicate this discussion here. The results are as follows. We decompose the moment as
\begin{equation}
\tilde{S}_{\rm M}^{jk} = \sum_{m=-2}^2 \tilde{S}^{jk}_m,
\label{SM_decomp}
\end{equation}
and find that each $m$-component is given by
\begin{equation}
\tilde{S}^{jk}_m = -\frac{M R^4}{c^2} \tilde{k}^m \tilde{B}^m_{jk},
\label{SM_m} 
\end{equation}
where $\tilde{B}^m_{jk}$ is the projection of the tidal moment defined by Eq.~(\ref{B_decomposition2}), and $\tilde{k}^m$ is a Love number specific to each value of $m$. They are given by
\begin{subequations}
\label{k_M} 
\begin{align}
\tilde{k}^{\pm 2}(\omega) &= -\frac{6}{5} \frac{2 \mp \omega/\Omega}{4 \mp 3\omega/\Omega}
\int_0^1 \hat{\rho}\, \hat{r}^6\, d\hat{r}, 
\label{k_m2} \\
\tilde{k}^{\pm 1}(\omega) &= \mp \frac{6}{\sqrt{30\pi}} (1 \mp \omega/\Omega)
\int_0^1 \hat{\rho}\, C_2^{\pm 1}(\omega,r) \hat{r}^3\, d\hat{r}, \\
\tilde{k}^{0}(\omega) &= \frac{1}{\sqrt{5\pi}} (\omega/\Omega)
\int_0^1 \hat{\rho}\, C_2^0(\omega,r) \hat{r}^3\, d\hat{r},
\label{k_m0} \\ 
\end{align}
\end{subequations}
in terms of the radial functions $C_2^m$ that appear in front of the $\ell = 2$ axial harmonics in Eq.~(\ref{va_decomp}). The Love numbers are real, and by virtue of Eq.~(\ref{C_reality}), they satisfy
\begin{equation}
\tilde{k}^{-m}(\omega) = \tilde{k}^m(-\omega).
\label{k_reality} 
\end{equation}
These results can be compared with those displayed in Sec.~V D of Ref.~\cite{poisson:20a}, which applied specifically to a tidal field produced by a companion body moving on a circular orbit. In this case the tidal moments are given by Eq.~(\ref{B_binary}), and the $J^{jk}$ of Eq.~(5.15) in Ref.~\cite{poisson:20a} agrees with the $S_{\rm M}^{jk}$ obtained by inserting Eq.~(\ref{SM}) within the Fourier integral. (Agreement requires summation over $m$; individual $m$-components do not correspond, because they are defined differently.) The comparison is aided by noting that $\tilde{k}^{\pm 2} = -2\gothj_{\pm 2}/3$, $\tilde{k}^{\pm 1} = \pm 2\gothj_{\pm 1}/3$, and $\tilde{k}^{0} = \gothj_{0}/3$, where $\gothj_m$ are the quantities defined by Eq.~(5.17) of Ref.~\cite{poisson:20a}.

\begin{table} 
\caption{\label{tab:overlaps} Eigenfrequencies $w_n^m := \omega_n^m/\Omega$, overlap integrals $\gothp_n^m$ with the gravitomagnetic tidal force density, and norms $\hat{N}_n^m$ of the dominant inertial modes, calculated for the stellar models described by Eq.~(\ref{rho_model}). These quantities are defined and computed in Sec.~VI of Ref.~\cite{poisson:20a}.}  
\begin{ruledtabular} 
\begin{tabular}{cccccc}
$b$ & $m$ & $n$ & $w_n^m$ & $\gothp_n^m$ & $\hat{N}_n^m$ \\ 
\hline 
  0 & 2 & $\bullet$ & $1.3333$ & $1.3244\e{-1}$ & $-2.1429\e{-1}$ \\
    & 1 & I & $1.1766$ & $1.3244\e{-1}$ & $9.2208\e{-2}$ \\
    & 1 & II & $-5.0994\e{-1}$ & $1.3244\e{-1}$ & $4.1970$ \\
    & 0 & I &  $8.9443\e{-1}$ & $3.2440\e{-1}$ & $8.5714\e{-1}$ \\
 & & & & & \\ 
  1 & 2 & $\bullet$ & $1.3333$ & $6.3502\e{-2}$ & $-1.0275\e{-1}$ \\
    & 1 & I & $1.4014$ & $-4.1974\e{-2}$ & $4.3258\e{-2}$ \\
    & 1 & II & $-4.1300\e{-1}$ & $2.6417\e{-2}$ & $3.0131\e{-1}$ \\
    & 0 & I & $1.0282$ & $-7.9947\e{-2}$ & $1.0879\e{-1}$ \\
 & & & & & \\ 
  2 & 2 & $\bullet$ & $1.3333$ & $3.5252\e{-2}$ & $-5.7039\e{-2}$ \\
    & 1 & I & $1.4979$ & $-3.1348\e{-2}$ & $5.3785\e{-2}$ \\
    & 1 & II & $-3.8059\e{-1}$ & $2.2777\e{-2}$ & $3.9049\e{-1}$ \\
    & 0 & I & $1.0763$ & $-6.3472\e{-2}$ & $1.2303\e{-1}$ \\
 & & & & & \\ 
  3 & 2 & $\bullet$ & $1.3333$ & $2.1933\e{-2}$ & $-3.5489\e{-2}$ \\
    & 1 & I & $1.5466$ & $-2.4483\e{-2}$ & $5.7869\e{-2}$ \\
    & 1 & II & $-3.6583\e{-1}$ & $1.9210\e{-2}$ & $4.4182\e{-1}$ \\
    & 0 & I & $1.1003$ & $-5.2790\e{-2}$ & $1.1369\e{-1}$
\end{tabular} 
\end{ruledtabular} 
\end{table} 

The perturbation created by the gravitomagnetic tidal field can also be represented as a sum over the body's normal modes of vibration. As shown in Ref.~\cite{poisson:20a}, the sum implicates only the body's inertial modes. And while in principle the sum should include an infinite number of terms, an excellent approximation results when it is truncated to only four modes. The relevant modes are those that have large overlap integrals with the tidal force density; they consist of an $m = 2$ $r$-mode labeled $(2, \bullet)$, two modes with $m=1$, one with positive frequency $(1, \mbox{I})$, another with negative frequency $(1, \mbox{II})$, and a final mode with $m=0$ labeled $(0, \mbox{I})$ --- actually a complex-conjugate pair of modes. The mode frequencies $w^m_n := \omega^m_n/\Omega$, overlap integrals $\gothp^m_n$, and norms $\hat{N}^m_n$ are listed in Table~\ref{tab:overlaps} (abridged from Ref.~\cite{poisson:20a}) for selected stellar models described by the mass density
\begin{equation}
\rho = C \biggl[ \frac{\sin(\pi r /R)}{(\pi r /R)} \biggr]^b,
\label{rho_model}
\end{equation}
where $C$ is a constant and $b$ an integer in the set $b = \{ 0, 1, 2, 3 \}$. With $b=0$ we have a body of constant density, and $b = 1$ gives the density of a polytrope with $p = K\rho^2$. As $b$ keeps on growing to $b=2$ and $b=3$, the body becomes increasingly centrally dense; these values of $b$ do not correspond to a recognizable equation of state.   

\begin{figure} 
\includegraphics[width=0.6\linewidth]{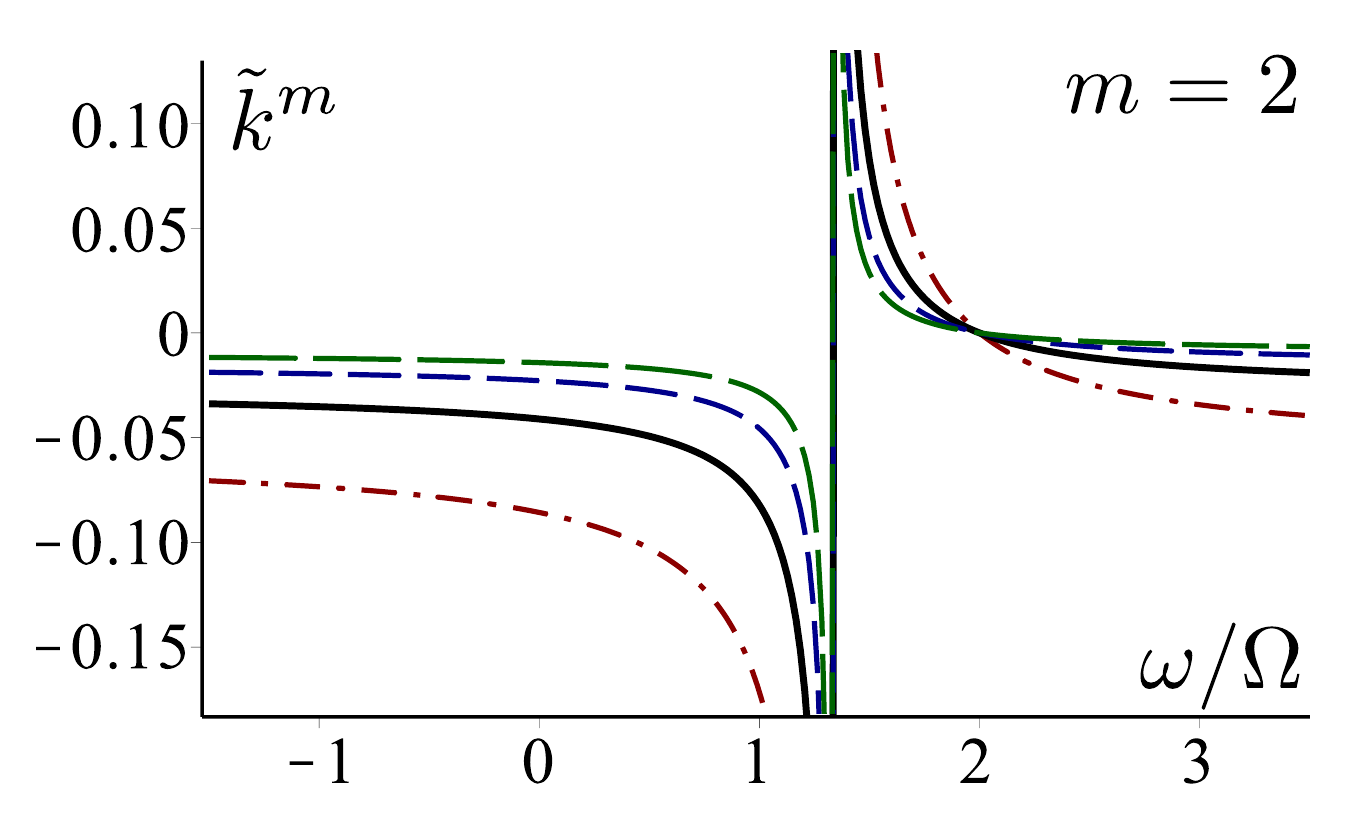}
\caption{Love number $\tilde{k}^m$ for $m=2$, plotted as a function of $\omega/\Omega$, computed for the density models of Eq.~(\ref{rho_model}). Red dash-dotted line: $b=0$. Black solid line: $b=1$. Blue short-dashed line: $b=2$. Green long-dashed line: $b=3$. The relevant inertial mode for $m=2$ is an $r$-mode, whose eigenfrequency $(4/3)\Omega$ is independent of the body's density profile. The resonance is indicated by the vertical line.}   
\label{fig:love_m2} 
\end{figure}

\begin{figure} 
\includegraphics[width=0.6\linewidth]{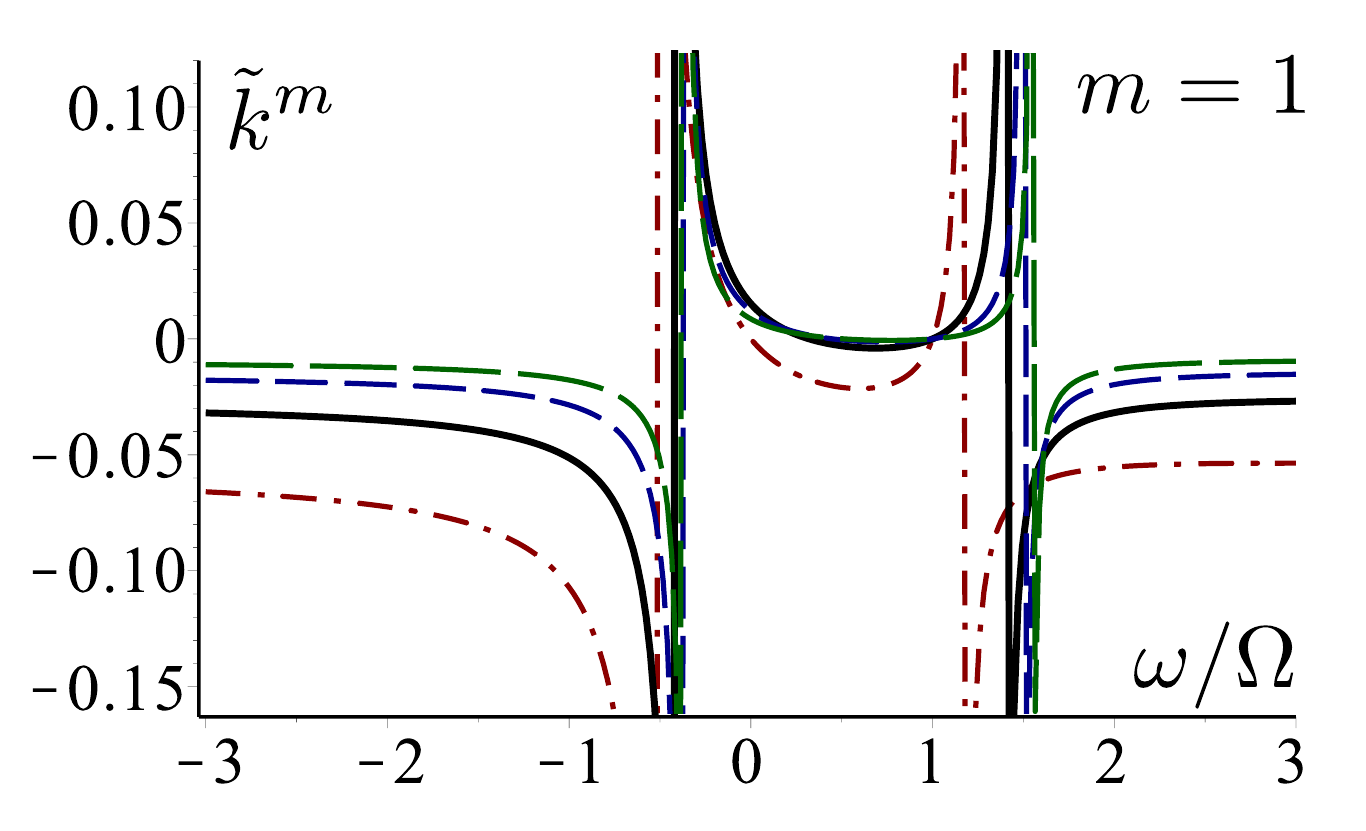}
\caption{Love number $\tilde{k}^m$ for $m=1$, plotted as a function of $\omega/\Omega$, computed for the density models of Eq.~(\ref{rho_model}). Red dash-dotted line: $b=0$. Black solid line: $b=1$. Blue short-dashed line: $b=2$. Green long-dashed line: $b=3$. There are two relevant inertial modes for $m=1$, one with positive eigenfrequency, the other with negative eigenfrequency. The resonances are indicated by vertical lines.}   
\label{fig:love_m1} 
\end{figure}

\begin{figure} 
\includegraphics[width=0.6\linewidth]{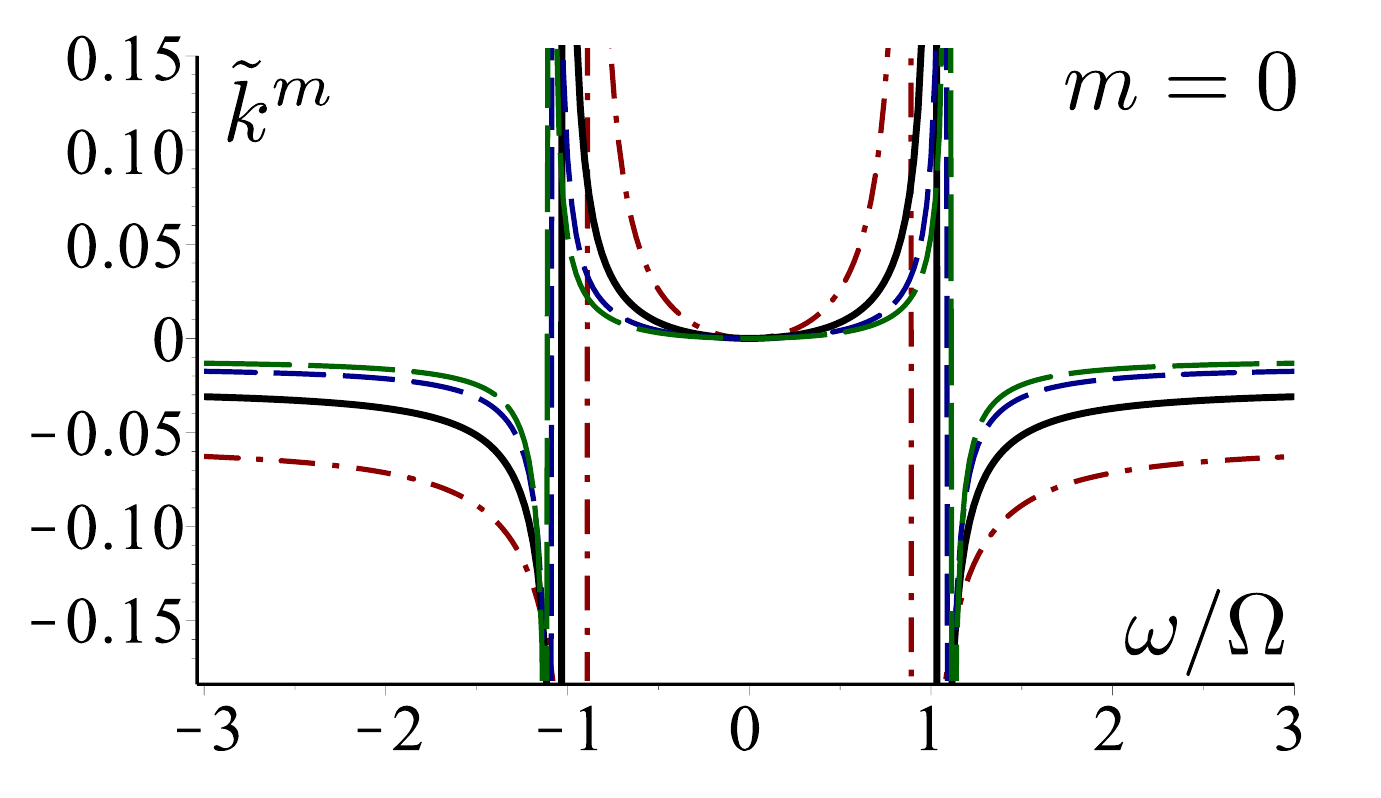}
\caption{Love number $\tilde{k}^m$ for $m=0$, plotted as a function of $\omega/\Omega$, computed for the density models of Eq.~(\ref{rho_model}). Red dash-dotted line: $b=0$. Black solid line: $b=1$. Blue short-dashed line: $b=2$. Green long-dashed line: $b=3$. The relevant inertial modes for $m=0$ are a pair of complex-conjugate modes with positive and negative eigenfrequencies of equal magnitude. The resonances are indicated by vertical lines.}   
\label{fig:love_m0} 
\end{figure}

The mode-sum representation of the Love numbers is given by 
\begin{subequations} 
\label{k_modesum} 
\begin{align} 
\tilde{k}^{\pm 2}(\omega) &= \frac{4\pi}{9} 
\frac{(\gothp_\bullet^2)^2}{\hat{N}_\bullet^2} \frac{2-w_\bullet^2}{w_\bullet^2} 
\frac{2 \mp w}{w_\bullet^2 \mp w}, 
\label{k_m2_modesum} \\
\tilde{k}^{\pm 1}(\omega) &= \sum_{n = {\rm I}, {\rm II}} \frac{4\pi}{9} 
\frac{(\gothp_n^1)^2}{\hat{N}_n^1} \frac{1-w_n^1}{w_n^1} 
\frac{1 \mp w}{w_n^1 \mp w}, \\  
\tilde{k}^{0}(\omega) &= \frac{4\pi}{27} 
\frac{(\gothp_{\rm I}^0)^2}{\hat{N}_{\rm I}^0} 
\frac{w^2}{(w_{\rm I}^0-w) (w_{\rm I}^0+w)},  
\end{align} 
\end{subequations} 
where $w := \omega/\Omega$. These equations can be compared to Eq.~(6.46) of Ref.~\cite{poisson:20a}. With $w^2_\bullet = 4/3$ and [Eqs.~(6.28) and (6.32) of Ref.~\cite{poisson:20a}, respectively]
\begin{equation} 
p^2_\bullet = 3\sqrt{\frac{3}{10\pi}}\int_0^1 \hat{\rho}\, \hat{r}^6\, d\hat{r}, \qquad 
\hat{N}^2_\bullet = -\frac{3}{2} \int_0^1 \hat{\rho}\, \hat{r}^6\, d\hat{r},   
\end{equation} 
Eq~(\ref{k_m2_modesum}) agrees precisely with Eq.~(\ref{k_m2}); for $m=1$ and $m=0$ the expressions for $\tilde{k}^m$ in Eqs~(\ref{k_modesum}) are approximations. The Love numbers obtained from Eqs.~(\ref{k_modesum}) are plotted as functions of $\omega/\Omega$ in Figs.~\ref{fig:love_m2}, \ref{fig:love_m1}, and \ref{fig:love_m0} for the selected values of $b$. The plots reveal the resonances that occur when the frequency $\omega$ of the tidal field matches one of the mode's eigenfrequencies, which vary with the body's density profile.

In Table~\ref{tab:static} we list $k^{\rm F}$ and $\tilde{k}^m(\omega = 0)$ for the selected values of $b$; here the computations are based on Eqs.~(\ref{k_F}) and (\ref{k_M}). We can see that as expected, the Love numbers decrease (in magnitude) as the body becomes centrally dense. An exception arises for the uniform-density model with $b=0$, for which $\tilde{k}^1(\omega = 0) = 0$. This happens because the solution to the perturbation equations for $m=1$ and $\rho = \mbox{constant}$ is 
\begin{equation} 
A_1^1(\omega=0,r) = \frac{1}{3} \sqrt{6\pi}\, \frac{r(r^2-R^2)}{R^3}, \qquad 
B_1^1(\omega=0,r) = \frac{1}{3} \sqrt{6\pi}\, \frac{r(2r^2-R^2)}{R^3}, \qquad 
C_2^1(\omega=0,r) = 0,   
\end{equation}  
with all other radial functions vanishing. The fact that the Love numbers with $m=0$ all vanish when $\omega = 0$ is a consequence of the factor of $\omega/\Omega$ in Eq.~(\ref{k_m0}). Finally, we note that by virtue of Eqs.~(\ref{k_F}) and (\ref{k_m2}), $\tilde{k}^2(\omega = 0) = -2k^{\rm F}$. 

\begin{table} 
\caption{\label{tab:static}  Field contribution $k^{\rm F}$ to the Love tensor, and matter contributions $\tilde{k}^m$ for $\omega = 0$, computed for the density models of Eq.~(\ref{rho_model}).}  
\begin{ruledtabular} 
\begin{tabular}{ccccc}
$b$ & $k^{\rm F}$ & $\tilde{k}^2(\omega=0)$ & $\tilde{k}^1(\omega=0)$ & $\tilde{k}^0(\omega=0)$ \\  
\hline 
0 & $4.2857\e{-2}$ & $-8.5714\e{-2}$ & $0.0000$ & $0.0000$ \\
1 & $2.0549\e{-2}$ & $-4.1099\e{-2}$ & $1.5236\e{-2}$ & $0.0000$ \\
2 & $1.1408\e{-2}$ & $-2.2816\e{-2}$ & $1.2035\e{-2}$ & $0.0000$ \\
3 & $7.0977\e{-3}$ & $-1.4195\e{-2}$ & $8.6028\e{-3}$ & $0.0000$ 
\end{tabular}
\end{ruledtabular}
\end{table} 
 
\section{Gravitomagnetic Love tensor of a slowly rotating body} 
\label{sec:tensor} 

We may now combine the results of Secs.~\ref{sec:field} and \ref{sec:matter} and construct the complete Love tensor of a slowly rotating body subjected to a gravitomagnetic tidal field. This includes the field contribution of Eq.~(\ref{k_F}) and the matter contributions of Eqs.~(\ref{k_M}); the fact that these depend on $m$ implies that the Love quantity cannot be a scalar --- it must be a tensor. 

We insert Eq.~(\ref{SM_m}) within Eq.~(\ref{SM_decomp}) and make use of Eqs.~(\ref{B_decomposition2}) and (\ref{B_projections}). We obtain 
\begin{equation} 
\tilde{S}^{\rm M}_{jk}(\omega) = -\frac{MR^4}{c^2} \tilde{k}^{{\rm M}\ pq}_{\ jk}(\omega)  
\tilde{\cal B}_{pq}(\omega),  
\end{equation} 
where 
\begin{equation} 
\tilde{k}^{{\rm M}\ pq}_{\ jk}(\omega) := \frac{8\pi}{15} \sum_{m=-2}^2 \tilde{k}^m(\omega) 
(\bar{\scrpt Y}_2^{m})_{jk} ({\scrpt Y}_2^{m})^{pq} 
\label{k_matter} 
\end{equation} 
is the matter contribution to the Love tensor. To this we add Eq.~(\ref{kF_tensor}), and get 
\begin{equation} 
\tilde{k}^{\ \ pq}_{jk}(\omega) := k^{\rm F}\, \delta_j^{\ p} \delta_k^{\ q} 
+ \tilde{k}^{{\rm M}\ pq}_{\ jk}(\omega) 
\label{k_complete} 
\end{equation} 
for the complete Love tensor. We recall that $k^{\rm F}$ is independent of $\omega$. An alternative, geometric representation for the Love tensor is constructed in Appendix~\ref{sec:geometric}. 

The relation between the tidal moment and the complete current quadrupole moment --- the sum of field and matter contributions --- is given by 
\begin{equation} 
\tilde{S}_{jk}(\omega) = -\frac{MR^4}{c^2}\tilde{k}_{jk}^{\ \ pq}(\omega) \tilde{\cal B}_{pq}(\omega).  
\label{SvsB_complete} 
\end{equation} 
It might be tempting to rewrite this equation in the time domain as a convolution of time-domain Love tensor and tidal moment. Such an expression, however, would not be well defined, because the frequency-domain Love tensor fails to go to zero when $\omega \to \pm \infty$; its time-domain version is therefore undefined. The frequency-domain tidal moment, however, can be trusted to go to zero sufficiently fast at large frequencies, and the Fourier transform of the combined right-hand side of Eq.~(\ref{SvsB_complete}) is therefore well defined. The time-domain current quadrupole moment $S_{jk}(t)$, as given by Eq.~(\ref{S_vs_B_convolution}), is well defined.    

\begin{acknowledgments} 
I thank Philippe Landry for helpful comments on an earlier draft of this paper. This work was supported by the Natural Sciences and Engineering Research Council of Canada.  
\end{acknowledgments} 

\appendix

\section{Nonrotating limit}
\label{sec:norotation}

In this Appendix we elucidate the $\Omega \to 0$ limit of the velocity perturbation displayed in Eqs.~(\ref{va_decomp}), keeping $\omega$ fixed. We first construct the velocity perturbation of a nonrotating body, and then see how this is recovered on the basis of the general results of Sec.~\ref{sec:tidal_force}. We also compute the matter contribution to the Love tensor in this limit, and show that it collapses to a single Love number. 

The perturbed Euler equation (\ref{euler}) reduces to 
\begin{equation} 
\partial_t \delta v_j + \partial_j (\delta h - \delta U) = \frac{4}{c^2} \partial_t U_j
\label{dE_norotation} 
\end{equation} 
when the body is nonrotating. A particular solution is
\begin{equation} 
\delta v_j = \frac{4}{c^2} U_j, 
\label{dv_nonrotating} 
\end{equation} 
together with $\delta h = \delta U = 0$. The general solution to the equation would also include a free oscillation described by a superposition of the body's normal modes; we eliminate this from the solution by choice of initial conditions. The velocity field of Eq.~(\ref{dv_nonrotating}) arises as a consequence of the relativistic circulation theorem \cite{shapiro:96, favata:06}. 

With $U_j$ given by Eq.~(\ref{Uj_tidal}), we have that  
\begin{equation} 
v_j = -\frac{2}{3c^2} \epsilon_{jkp} {\cal B}^p_{\ q} x^k x^q.  
\end{equation} 
Transforming to spherical coordinates, making use of the decomposition of Eqs.~(\ref{B_decomposition1}), (\ref{B_decomposition2}), and recalling the definition (\ref{axial}) of the axial harmonics, this is
\begin{equation}
\delta v^a = \frac{1}{3c^2} r^3 \sum_{m=-2}^2 {\cal B}^m (X_2^m)^a.
\label{va_norotation} 
\end{equation}
This expression reveals that the velocity field is a pure quadrupole of axial parity.  

We wish to figure out how Eqs.~(\ref{va_decomp}) reduce to Eq.~(\ref{va_norotation}) when $\Omega \to 0$. We observe immediately that the $m = \pm 2$ terms match when $w := \omega/\Omega \to \infty$. For $m=\pm 1$ and $m=0$, however, we must dig deep into the explicit form of the fluid equations, and work out what happens when $w \to \infty$.

We require the curl of Euler's equation, 
\begin{equation}
Z^a := \epsilon^{abc} \nabla_b \delta \tilde{E}_c - \tilde{q}^a,
\label{curl} 
\end{equation}
where
\begin{equation}
\delta \tilde{E}_a := -i\omega\delta \tilde{v}_a + \delta \tilde{v}^b\, \nabla_b v_a
+ v^b \nabla_b \delta \tilde{v}_a + \nabla_a (\delta \tilde{h} - \delta \tilde{U})
\end{equation}
is the perturbed Euler operator, and $\tilde{q}^a$ is the curl of the gravitomagnetic tidal force density, as given by Eq.~(\ref{q_decomp}); $v^a$ is the rotational velocity of the unperturbed body. The gradient term in $\delta \tilde{E}_a$ is eliminated by taking the curl, and Eq.~(\ref{curl}) involves $\delta \tilde{v}_a$ only. Because $\nabla_a Z^a = 0$, Eq.~(\ref{curl}) supplies us with only two independent equations. A third is provided by the continuity equation $\nabla_a (\rho \delta \tilde{v}^a) = 0$, which gives rise to
\begin{equation}
r \frac{dA_\ell^m}{dr} + \biggl( 1 + \frac{r}{\rho} \frac{d\rho}{dr} \biggr) A_\ell^m
- \ell(\ell+1) B_\ell^m = 0.
\label{continuity}
\end{equation}
The surface condition $r_a \delta \tilde{v}^a = 0$ at $r = R$ produces $A_\ell^m(r=R) = 0$.

For $m=1$, $Z^r = 0$ yields (we omit the $m$ labels on the radial functions and spherical harmonics, to avoid a clutter of notation) 
\begin{align}
0 &= \sum_{\ell} \Bigl\{ 2(\sin\theta\, \partial_\theta + 2\cos\theta) A_\ell\, Y_\ell
- 2\bigl[ \sin\theta\, \partial_\theta + \ell(\ell+1) \cos\theta \bigr] B_\ell\, Y_\ell
+ \bigl[ \ell(\ell+1)(w-1) + 2 \bigr] C_\ell\, Y_\ell \Bigr\}
\nonumber \\ & \quad \mbox{} 
- 3 (r/R)^3 \sin\theta\cos\theta\, e^{i\phi}.
\end{align}
From $Z^\theta = 0$ we get
\begin{align}
0 &= \sum_{\ell} \Bigl\{ 2\sin^2\theta\, r A_\ell'\, Y_\ell
+ \bigl[ 2\sin\theta\cos\theta\, \partial_\theta + w - 1 \bigr] r B_\ell'\, Y_\ell
- \bigl[ (w-1)\sin\theta\, \partial_\theta + 2\cos\theta \bigr] r C_\ell'\, Y_\ell
\nonumber \\ & \quad \mbox{} 
+ (w-1) A_\ell\, Y_\ell - 2 B_\ell\, Y_\ell + 2\sin\theta\, C_\ell\, \partial_\theta Y_\ell \Bigr\} 
+ \frac{3}{2} (r/R)^3 \sin\theta (2\cos^2\theta - 1) e^{i\phi}, 
\end{align}
with a prime indicating differentiation with respect $r$. For $m=0$, $Z^r = 0$ gives
\begin{equation}
0 = \sum_{\ell} \Bigl\{ 2(\sin\theta\, \partial_\theta + 2\cos\theta) A_\ell\, Y_\ell
- 2\bigl[ \sin\theta\, \partial_\theta + \ell(\ell+1) \cos\theta \bigr] B_\ell\, Y_\ell
- \ell(\ell+1) w C_\ell\, Y_\ell \Bigr\}
- 3 (r/R)^3 (3\cos^2\theta - 1), 
\end{equation}
and $Z^\phi = 0$ produces
\begin{equation}
0 = \sum_{\ell} \Bigl\{ w r B_\ell'\, \partial_\theta Y_\ell
+ 2\cos\theta\, r C_\ell'\, \partial_\theta Y_\ell
- w A_\ell\, \partial_\theta Y_\ell
+ 2\ell(\ell+1) \sin\theta\, C_\ell\, Y_\ell \Bigr\}.
\end{equation}

We insert a small-$\Omega$ expansion of the form
\begin{equation}
A_\ell = \mbox{}_0 A_\ell + w^{-1} \mbox{}_1 A_\ell + O(w^{-2}), \qquad
B_\ell = \mbox{}_0 B_\ell + w^{-1} \mbox{}_1 B_\ell + O(w^{-2}), \qquad
C_\ell = \mbox{}_0 C_\ell + w^{-1} \mbox{}_1 C_\ell + O(w^{-2})
\end{equation}
within the perturbation equations, and collect terms of equal order in $w$. From $Z^a = 0$ at order $w$ we obtain
\begin{equation}
\mbox{}_0 C_\ell = 0, \qquad r\, \mbox{}_0 B_\ell' - \mbox{}_0 A_\ell = 0. 
\end{equation}
The second equation is substituted within Eq.~(\ref{continuity}), and we find that $\mbox{}_0 B_\ell$ must satisfy the differential equation
\begin{equation}
r^2\, \mbox{}_0 B_\ell'' + \biggl(2 + \frac{r}{\rho} \frac{d\rho}{dr} \biggr) r\, \mbox{}_0 B_\ell'
- \ell(\ell+1)\, \mbox{}_0 B_\ell = 0.
\end{equation} 
The general solution comes with two constants of integration. One must be chosen to eliminate the $r^{-(\ell+1)}$ behavior near $r = 0$, and the other remains as an overall multiplicative factor. This leaves us with insufficient freedom to impose the boundary condition of $A_\ell(r=R)=0$, and we conclude that the only acceptable solution is $\mbox{}_0 B_\ell = 0$, which implies $\mbox{}_0 A_\ell = 0$.

At order $w^0$ we find that $Z^r = 0$ yields
\begin{equation}
\sum_\ell \ell(\ell+1)\, \mbox{}_1 C_\ell\, Y_\ell = 3 (r/R)^3 \sin\theta\cos\theta\, e^{i\phi}
\end{equation}
for $m=1$, and
\begin{equation}
\sum_\ell \ell(\ell+1)\, \mbox{}_1 C_\ell\, Y_\ell = -3 (r/R)^3 (3\cos^2\theta-1) 
\end{equation}
for $m=0$. It follows that
\begin{equation}
\mbox{}_1 C_2^1 = -\frac{1}{15} \sqrt{30\pi} (r/R)^3, \qquad
\mbox{}_1 C_2^0 = -\frac{2}{5} \sqrt{5\pi} (r/R)^3, 
\end{equation}
and $\mbox{}_1 C_\ell =0$ for all other values of $\ell$. The angular equations at order $w^0$ then reveal that 
\begin{equation}
r\, \mbox{}_1 B_\ell' - \mbox{}_1 A_\ell = 0.
\end{equation}
Inserting this within the continuity equation, we find once again that the only acceptable solution is $\mbox{}_1 A_\ell = \mbox{}_1 B_\ell = 0$.

With
\begin{equation}
C_2^1 = -\frac{1}{w} \frac{1}{15} \sqrt{30\pi} (r/R)^3 + O(w^{-2}), \qquad
C_2^0 = -\frac{1}{w} \frac{2}{5} \sqrt{5\pi} (r/R)^3 + O(w^{-2}) 
\end{equation}
as the only nonvanishing radial functions, we see that Eq.~(\ref{va_decomp}) does indeed give rise to the velocity field of Eq.~(\ref{va_norotation}) in the $\Omega \to 0$ limit.

To conclude the discussion we compute the current quadrupole moment that results from the velocity perturbation of Eq.~(\ref{va_norotation}). Inserting Eq.~(\ref{va_norotation}) within Eq.~(\ref{SM}) and performing the angular integration, we find that the current quadrupole moment admits the same decomposition as in Eq.~(\ref{SM_decomp}), with $m$-components still given by Eq.~(\ref{SM_m}). The difference is that the Love numbers are now independent of $m$, and given by 
\begin{equation}
\tilde{k}^m = k^{\rm M} := -\frac{2}{5} \int_0^1 \hat{\rho}\, \hat{r}^6\, d\hat{r}.
\label{k_norotation} 
\end{equation}
The sum over $m$ can then be carried out, and we obtain
\begin{equation}
S^{jk}_{\rm M} = -\frac{MR^4}{c^2} k^{\rm M} {\cal B}^{jk};
\end{equation}
because the Love numbers no longer depend on $\omega$, the relation can be formulated directly in the time domain. We see that the independence of the Love numbers on $m$ turns the Love tensor of Sec.~\ref{sec:tensor} into a single Love number. The result of Eq.~(\ref{k_norotation}) agrees with Eq.~(6.20) of Ref.~\cite{landry-poisson:15b} --- it is the coefficient of the $\lambda$ term when $\ell = 2$. 

\section{Multipole expansion of the vector potential}
\label{sec:anapole} 

In this Appendix we construct the multipole expansion of the matter contribution to the vector potential. According to Eq.~(\ref{UM_poisson}), this is a solution to
\begin{equation}
\nabla^2 U^j = -4\pi G J^j,
\label{Uj_poisson_rep} 
\end{equation} 
where $J^j := \rho \delta v^j$ is the current density associated with the velocity perturbation. Here and below 
we omit the label ``M'' on the vector potential; it is clear that we are focusing our attention entirely on the matter contribution. 

By virtue of the slow-rotation approximation for the fluid perturbation, the current density satisfies $\partial_j J^j = 0$; refer to the discussion that leads to Eq.~(\ref{continuity}). We use this property to derive a number of integral relations satisfied by the current density. First, the identity $\partial_k( J^k x^j) = J^j$ implies
\begin{equation}
\int J^j\, dV = 0
\label{identity1}
\end{equation}
after converting the volume integral into a surface integral and noting that $J^k r_k = 0$ on the surface.
Similarly, integration of $\partial_p(J^p x^j x^k)$ yields 
\begin{equation}
\int \bigl( J^j x^k + x^j J^k \bigr)\, dV = 0.  
\label{identity2}
\end{equation} 
Finally, integration of $\partial_q(J^q x^j x^k x^p)$ gives
\begin{equation}
\int \bigl( J^j x^k x^p + x^j J^k x^p + x^j x^k J^p \bigr)\, dV = 0, 
\label{identity4}
\end{equation}
from which we also get
\begin{equation} 
\int \bigl( J^j r^2 + 2 x^j\, J_k x^k \bigr)\, dV = 0.
\label{identity5} 
\end{equation}

Next we decompose
\begin{equation} 
A^{jk} := \int J^j x^k\, dV, \qquad
B^{jkp} := \int J^j x^k x^p\, dV
\end{equation} 
into irreducible components. An arbitrary tensor $A_{jk}$ can always be decomposed into symmetric and antisymmetric pieces, and the latter can always be represented by a vector $A^j$. We write
\begin{equation}
A_{jk} = A_{(jk)} + \epsilon_{jkp} A^p,
\end{equation}
with $A^j := \frac{1}{2} \epsilon^{jkp} A_{kp}$. In our case, the symmetric part of $A_{jk}$ vanishes by virtue of Eq.~(\ref{identity2}), and up to a numerical factor, $A^j$ is recognized as the angular-momentum vector associated with the current density. We have obtained
\begin{equation}
\int J^j x^k\, dV = -\frac{1}{2} \epsilon^{jkp} S_p, \qquad
S^j := \epsilon^j_{\ kp} \int x^k J^p\, dV,
\label{one_moment} 
\end{equation}
where $S^j$ is the angular momentum contributed by the velocity perturbation.

A tensor $B_{jkp}$, symmetric in the last two indices, is decomposed as (see Appendix A of Ref.~\cite{poisson-corrigan:18}) 
\begin{equation}
B_{j(kp)} = B_\stf{jkp} + \epsilon_{jk}^{\ \ q} B_\stf{qp} + \epsilon_{jp}^{\ \ q} B_\stf{qk}
+ \delta_{jk} (\bar{B}_p - B_p) + \delta_{jp} (\bar{B}_k - B_k)
+ \delta_{kp} (\bar{B}_j + 2 B_j),
\end{equation}
where $B_\stf{jkp}$ is the symmetric-tracefree piece of the original tensor, and
\begin{equation}
B_\stf{jk} := \frac{1}{3} \epsilon_{pq(j} B^{pq}_{\ \ k)}, \qquad
\bar{B}_j := \frac{1}{15}(B_{jk}^{\ \ k} + 2 B^k_{\ kj}), \qquad
B_j := \frac{1}{6} (B_{jk}^{\ \ k} - B^k_{\ kj}).
\end{equation}
In our case we find that $B_{(jkp)} = 0$ by virtue of Eq.~(\ref{identity4}), which implies that $B_\stf{jkp} = 0$. We also have that $\bar{B}_j = 0$ follows from Eq.~(\ref{identity5}). From this we get that $B_j$ can be reduced to $\int J_k r^2\, dV$, up to a numerical factor. Finally, $B_\stf{jk}$ is recognized as the current quadrupole moment, also up to a numerical factor. We have obtained
\begin{equation} 
\int J^j x^k x^p\, dV = -\frac{1}{3} \bigl( \epsilon^{jkq} S_q^{\ p} + \epsilon^{jpq} S_q^{\ k} \bigr) 
- \frac{1}{4} \bigl( \delta^{jk} K^p + \delta^{jp} K^k \bigr) + \frac{1}{2} \delta^{kp} K^j,
\label{two_moment} 
\end{equation}
where
\begin{equation}
S^{jk} := \epsilon^{(j}_{\ \ pq} \int x^{k)} x^p J^q\, dV
\label{current quadrupole} 
\end{equation}
is the current quadrupole moment, and
\begin{equation}
K^j := \int J^j r^2\, dV
\label{anapole} 
\end{equation}
is the so-called anapole moment. 

The solution to Eq.~(\ref{Uj_poisson_rep}) is 
\begin{equation}
U^j(\bm{x}) = G \int \frac{J^j(\bm{x'})}{|\bm{x}-\bm{x'}|}\, dV',
\end{equation}
where $\bm{x'}$ is a point inside the body, and $dV'$ is the volume element surrounding it. We take $\bm{x}$ to be outside the body, and we expand the vector potential in inverse powers of $r := |\bm{x}|$. We write
\begin{equation}
\frac{1}{|\bm{x}-\bm{x'}|} = \frac{1}{r} - x^{\prime k} \partial_k \frac{1}{r}
+ \frac{1}{2} x^{\prime k} x^{\prime p} \partial_{kp} \frac{1}{r} + O(r^{-4}),
\end{equation}
insert this within the integral, and make use of Eqs.~(\ref{identity1}), (\ref{one_moment}), and (\ref{two_moment}). We arrive at
\begin{equation}
U^j = -\frac{1}{2} G \epsilon^j_{\ kp} S^k\, \partial^p \frac{1}{r}
- \frac{1}{3} G \epsilon^j_{\ kp} S^p_{\ q}\, \partial^{kq} \frac{1}{r}
- \frac{1}{4} G K_k\, \partial^{jk} \frac{1}{r} + O(r^{-4}).
\end{equation}
This expression makes it clear that the anapole term in $U^j$ is a pure gradient. It can always be removed with the gauge transformation
\begin{equation}
U^j_{\rm new} = U^j_{\rm old} + \partial^j f, \qquad
f := \frac{1}{4} G K^k\, \partial_k \frac{1}{r}.
\end{equation}
This term, therefore, plays no role in the post-Newtonian tidal interaction of a slowly rotating body.

An alternative expression for the vector potential is obtained by evaluating the derivatives of $r^{-1}$. We have
$\partial_j r^{-1} = -r_j/r^2$ and $\partial_{jk} r^{-1} = 3r_\stf{jk}/r^3$, where $r_j := \partial_j r$ is the unit radial vector and $r_\stf{jk} := r_j r_k - \frac{1}{3} \delta_{jk}$. We find that the vector potential becomes
\begin{equation}
U^j = \frac{1}{2} G \epsilon^j_{\ kp} S^k\, \frac{r^p}{r^2} 
- G \epsilon^j_{\ kp} S^p_{\ q}\, \frac{r^k r^q}{r^3} 
- \frac{3}{4} G K_k\, \frac{r^\stf{jk}}{r^3} + O(r^{-4}).
\end{equation}
The current quadrupole moment was computed in Sec.~\ref{sec:matter}. In the remainder of this Appendix we show that $S^j = 0$ --- the velocity perturbation makes no contribution to the body's angular momentum --- and calculate the anapole moment.

To compute $S^j$ we insert the velocity field of Eqs.~(\ref{va_decomp}) within the definition of Eq.~(\ref{one_moment}). The angular integrations are evaluated with
\begin{subequations}
\label{angular1} 
\begin{align}
\int \epsilon^j_{\ kp} r^k r^p\, Y_\ell^m\, d\Omega &=0,
\label{ang_a} \\
\int \epsilon^j_{\ kp} r^k (Y_\ell^m)^p\, d\Omega &=0,
\label{ang_b} \\
\int \epsilon^j_{\ kp} r^k (X_\ell^m)^p\, d\Omega &=
\frac{8\pi}{3r} ( \bar{\scrpt Y}_1^m )^j\, \delta_{\ell,1}, 
\label{ang_c} 
\end{align}
\end{subequations}
where the constant vector on the right of Eq.~(\ref{ang_c}) is such that 
\begin{equation} 
Y_1^m(\theta,\phi) =  (\bar{\scrpt Y}_1^{m})_j\, r^j;  
\end{equation} 
it is the $\ell = 1$ analogue of the tensors introduced in Eq.~(\ref{STFtensors}) for $\ell = 2$. The identity of Eq.~(\ref{ang_a}) follows directly from the antisymmetry of the permutation symbol and the symmetry of $r^k r^p$; the spherical harmonics and the integration play no role. The result of Eq.~(\ref{ang_b}) is obtained after integration by parts. And we get Eq.~(\ref{ang_c}) by combining the permutation symbols, integrating by parts, expressing $r^j$ in terms of $\ell = 1$ spherical harmonics, and invoking the orthonormality of spherical harmonics. Equations~(\ref{angular1}) reveal that a nonvanishing angular-momentum vector would come from an eventual $C_1^m(\omega,r) \neq 0$ in the velocity perturbation; but as Eqs.~(\ref{va_decomp}) indicates, these radial functions vanish. We have found that indeed, the velocity perturbation makes no contribution to the body's angular momentum.

To calculate the anapole moment of Eq.~(\ref{anapole}), we make use of the angular integrals
\begin{subequations}
\label{angular2} 
\begin{align}
\int r^j\, Y_\ell^m\, d\Omega &= \frac{4\pi}{3} (\bar{\scrpt Y}_1^{m})_j\, \delta_{\ell,1}, \\
\int (Y_\ell^m)^j\, d\Omega &=\frac{8\pi}{3r} (\bar{\scrpt Y}_1^{m})_j\, \delta_{\ell,1}, \\
\int (X_\ell^m)^j\, d\Omega &= 0, 
\end{align}
\end{subequations}
which are obtained by straightforward manipulations, along the lines required to establish Eqs.~(\ref{angular1}).  These results imply that $K^j$ comes from $\ell = 1$ contributions to the velocity perturbation; it implicates the radial functions $A_1^m(\omega,r)$, $B_1^m(\omega,r)$ with $m = \pm 1$ and $m = 0$. After inserting Eqs.~(\ref{va_decomp}) within Eq.~(\ref{anapole}), we get
\begin{equation}
\tilde{K}^j = \sum_{m=-1}^1 \tilde{K}^j_m
\end{equation}
with
\begin{equation}
\tilde{K}^j_m = -\frac{4\pi i}{3} \alpha^m R^3 \frac{\tilde{\cal B}^m}{c^2} 
(\omega/\Omega - |m|) (\bar{\scrpt Y}_1^{m})^j \int \rho r^3 (A_1^m + 2 B_1^m)\, dr,
\end{equation}
where $\alpha^{\pm 1} = \mp 5/\sqrt{30\pi}$ and $\alpha^0 = 5/(6\sqrt{5\pi})$. The identity 
\begin{equation}
\int \rho r^3 (3 A_1^m + 2 B_1^m)\, dr = 0,
\end{equation}
which follows from Eq.~(\ref{continuity}) with $\ell = 1$, can be used to simplify the radial integration. We may choose, for example, to eliminate $B_1^m$ in favor of $A_1^m$, or vice-versa. 

\section{Geometric representation for the Love tensor}
\label{sec:geometric}

The matter contribution to the Love tensor is given by Eq.~(\ref{k_matter}), which we re-express as
\begin{equation}
\tilde{k}^{\rm M}_{jkpq}(\omega) = \sum_{m=-2}^2 \tilde{k}^m(\omega)\, {\cal P}^m_{jkpq},
\label{k_start} 
\end{equation}
where
\begin{equation}
{\cal P}^m_{jkpq} := \frac{8\pi}{15} (\bar{\scrpt Y}_2^{m})_{jk}({\scrpt Y}_2^{m})_{pq}.
\end{equation}
We recall from Eq.~(\ref{k_reality}) that $\tilde{k}^{-m}(\omega) = \tilde{k}^m(-\omega)$, and we also have that
$({\scrpt Y}_2^{-m})^{jk} = (-1)^m (\bar{\scrpt Y}_2^{m})^{jk}$. In this Appendix we wish to express the Love tensor in terms of geometric quantities that are intrinsic to the problem, namely the metric $\delta_{jk}$, the permutation symbol $\epsilon_{jkp}$, and the unit vector $e^j$ that points in the direction of the axis of rotation.

To achieve this we first extract from Eq.~(\ref{k_start}) the real and imaginary parts of the Love tensor. These are given by
\begin{subequations}
\label{k_re_im} 
\begin{align}
\mbox{Re}[\tilde{k}_{jkpq}] &= \frac{1}{2} \Bigl[ \tilde{k}^2(\omega) + \tilde{k}^2(-\omega) \Bigr] {\cal Q}^{+2}_{jkpq}
+ \frac{1}{2} \Bigl[ \tilde{k}^1(\omega) + \tilde{k}^1(-\omega) \Bigr] {\cal Q}^{+1}_{jkpq}
+ \tilde{k}^0(\omega)\, {\cal Q}^0_{jkpq}, \\
\mbox{Im}[\tilde{k}_{jkpq}] &= \frac{1}{2} \Bigl[ \tilde{k}^2(\omega) - \tilde{k}^2(-\omega) \Bigr] {\cal Q}^{-2}_{jkpq}
+ \frac{1}{2} \Bigl[ \tilde{k}^1(\omega) - \tilde{k}^1(-\omega) \Bigr] {\cal Q}^{-1}_{jkpq},
\end{align}
\end{subequations}
where
\begin{subequations}
\begin{align}
{\cal Q}^{+2}_{jkpq} &:= {\cal P}^2_{jkpq} + {\cal P}^{-2}_{jkpq}, \\
{\cal Q}^{-2}_{jkpq} &:= -i \bigl( {\cal P}^2_{jkpq} - {\cal P}^{-2}_{jkpq} \bigr), \\
{\cal Q}^{+1}_{jkpq} &:= {\cal P}^1_{jkpq} + {\cal P}^{-1}_{jkpq}, \\
{\cal Q}^{-1}_{jkpq} &:= -i \bigl( {\cal P}^1_{jkpq} + {\cal P}^{-1}_{jkpq} \bigr), \\
{\cal Q}^{0}_{jkpq} &:= {\cal P}^0_{jkpq}.
\end{align}
\end{subequations}
Next we introduce the vector basis $x^j = (1, 0, 0)$, $y^j = (0, 1, 0)$, and $z^j = (0, 0, 1)$ attached to the Cartesian coordinate system, and proceed to express ${\cal Q}^m_{jkpq}$ in terms of these vectors. We find that the tensors simplify into forms that involve the combinations $h_{jk} := x_j x_k + y_j y_k$ and $\epsilon_{jk} = x_j y_k - y_j x_k$, in addition to $z_j$. Explicitly, we have that
\begin{subequations}
\begin{align}
{\cal Q}^{+2}_{jkpq} &=\frac{1}{2} \bigl( -h_{jk} h_{pq} + h_{jp} h_{kq} + h_{jq} h_{kp} \bigr), \\
{\cal Q}^{-2}_{jkpq} &= -\frac{1}{4} \bigl( \epsilon_{jp} h_{kq} + \epsilon_{jq} h_{kp}
+ \epsilon_{kp} h_{jq} + \epsilon_{kq} h_{jp} \bigr), \\
{\cal Q}^{+1}_{jkpq} &= \frac{1}{2} \bigl( h_{jp} z_k z_q + h_{jq} z_k z_p
+ h_{kp} z_j z_q + h_{kq} z_j z_p \bigr), \\
{\cal Q}^{-1}_{jkpq} &= -\frac{1}{2} \bigl( \epsilon_{jp} z_k z_q + \epsilon_{jq} z_k z_p
+ \epsilon_{kp} z_j z_q + \epsilon_{kq} z_j z_p \bigr), \\
{\cal Q}^{0}_{jkpq} &= \frac{1}{6} \bigl( h_{jk} - 2 z_j z_k \bigr) \bigl( h_{pq} - 2 z_p z_q \bigr). 
\end{align}
\end{subequations}
The final step is to eliminate the dependence on the coordinate system by noting that $h_{jk}$, $\epsilon_{jk}$ and $z_j$ are geometric objects given by 
\begin{equation} 
h_{jk} = \delta_{jk} - e_j e_k, \qquad
\epsilon_{jk} = \epsilon_{jkp} e^p, \qquad
z_j = e_j.
\end{equation}
With this transcription, the projectors ${\cal Q}^m_{jkpq}$ are written in terms of intrinsic quantities. Making the substitution in Eq.~(\ref{k_re_im}), we obtain a geometric representation for the Love tensor. 

\bibliography{/Users/poisson/writing/papers/tex/bib/master}
\end{document}